\documentclass[multphys,vecphys]{svmult}

\usepackage{makeidx}         
\usepackage{graphicx}        
\usepackage{multicol}        
\usepackage[bottom]{footmisc}

\makeindex             

\begin{document}

\title*{The evolution of galaxy groups and of galaxies therein}
\titlerunning{Evolution of groups \& galaxies therein}
\author{Gary A. Mamon
}
\institute{IAP, 98 bis Bd Arago, F--75014, Paris, France,
\texttt{gam@iap.fr}}

%
%
\maketitle

\begin{abstract}
Properties of groups of galaxies depend sensitively on the algorithm for
group selection, and even the most recent catalogs of groups built from
redshift-space selection should suffer from projections and inclusion of
infalling galaxies.
The cosmo-dynamical evolution of groups from initial Hubble expansion to
collapse and virialization leads to a fundamental track in virial-theorem
estimated $M/L$ vs crossing time.
The increased rates of mergers, both direct and after orbital
decay by dynamical friction, in (low velocity dispersion) groups relative to
clusters, explain the higher fraction of elliptical galaxies at
given local number density in
X-ray selected groups, relative to clusters, even when the hierarchical
evolution of groups is considered.
Galaxies falling into groups and clusters should later travel outwards to
typically 2 virial radii, which is close to but somewhat less than 
the outermost radius where
galaxy star formation efficiencies are observed to be
enhanced relative to field galaxies of
same morphological type.
An ongoing analysis of the internal kinematics of X-ray selected groups
suggests that the radial profiles of line of sight velocity dispersion are
consistent with isotropic NFW distributions for the total mass density, with
higher concentrations in massive groups than $\Lambda$CDM predictions and lower
concentrations in low mass groups. The critical mass, at $M_{200} \approx
10^{13} M_\odot$ is consistent with 
possible breaks in the X-ray luminosity-temperature and Fundamental Plane
relations. The internal kinematics of groups indicate that the $M-T$ relation
of groups should agree with that extrapolated from clusters with no break at
the group scale.
The analyses of observed velocity dispersion profiles and of the fundamental
track both suggest that 
low velocity dispersion groups (compact and loose, X-ray emitting or
undetected) are quite
contaminated by chance projections.

\end{abstract}

\section{Introduction}
The attractive nature of gravity tends to assemble galaxies together in
groups.
With typical grouping algorithms, roughly half of all galaxies reside in
groups. A smaller fraction of galaxies live in virialized groups of at least
4 bright galaxies, and a considerably smaller fraction live in the more
massive virialized
clusters.


Given typical scaling relations, defining the virial radius of groups where
the mean density is 200 times the critical density of the Universe, groups of
galaxies have ranges of
mass within the virial radius, virial
and turnaround radii (all assuming $H_0 = 70 \,\rm km \,s^{-1} \, Mpc^{-1}$), 
velocity dispersion and temperature shown in Table~\ref{mamon:grouppars}.
\begin{table}[ht]
\centering
\caption{Typical scales of groups}
\begin{tabular}{lccccr}
\hline
  & $\log M_{200}$ & $r_{200}$ & $r_{\rm ta}$ & $\sigma_v$ & \multicolumn{1}{c}{$kT$} \\
  & ($M_\odot$) & (Mpc) & (Mpc) & ($\, \rm km \, s^{-1}$) & (keV) \\
\hline
Minimum & 12.5 & 0.3 &  1.1 & 140 & 0.2\ \ \ \\
Maximum & 14.0 & 1.0 & 3.4 & 450 & 2\ \ \  \\
\hline
\end{tabular}
\label{mamon:grouppars}
\end{table}
More massive objects can be called clusters. Of course, the limiting mass
between groups and clusters is arbitrary and historical. The more massive
groups have properties (e.g. $L_X-T$ \cite{mamon:OP04}
and Fundamental Plane \cite{mamon:SMCB93} relations) 
expected from the extrapolation of clusters,
  while the less massive groups do not appear to follow such extrapolations, 
with the separation between massive cluster-like and low mass groups
occurring at $M_{200} \approx 10^{13}\,M_\odot$.

Groups of galaxies thus provide an important laboratory to understand how the
density of the environment affects the properties of galaxies.
In turn, the modulation with environment of galaxy properties serves as an
important constraint for (semi-)analytical models of galaxy
formation. 

This review focusses on several dynamical and cosmological aspects of the
evolution of groups and of their constituent galaxies.

\section{The evolution of groups}
\subsection{Group expansion, collapse and virialization}
Groups are difficult to define from galaxy catalogs,
because the selection of nearby galaxies in redshift space causes frequent
interlopers and spurious groups 
\cite{mamon:MFW93,mamon:DKCW99} 
(also Eke, at this meeting). 
This problem of interlopers and spurious groups
is probably worsened by the use of the Friends of Friends grouping
algorithm, which tends to produce filamentary structures when there are less
than a dozen objects (it would be worthwhile to compare the efficiency of the
Friends-of-Friends algorithm with other grouping methods, based upon
cosmological simulations).


While virialized structures in a
$\Lambda$CDM Universe with $\Omega_m=0.24$ \cite{mamon:Spergel+06} 
have a mean density $\Delta=384$ \cite{mamon:KS96} times the mean density of the Universe,
i.e. 94 times the critical
density 
of the Universe, 
groups have been selected in the past with Friends-of-Friends
linking lengths corresponding to an overdensity of only 20 \cite{mamon:GH83} or 80
\cite{mamon:RGH89,mamon:MZ02,mamon:MZ05}. In comparison, the mean density at the turnaround
radius in a $\Lambda$CDM Universe with $\Omega_m = 0.24$ is 14 times that of the
Universe (Mamon, in preparation). Overdensities of 80 relative to the mean
Universe density thus roughly
correspond to the geometric mean between the virial and turnaround radii,
and thus, \emph{in most catalogs, group galaxies are selected out to the
  region of rapid infall, and the infalling galaxies  
bias the group definition, mass estimate and properties}.
Given very large galaxy samples, such as SDSS, group selection should be done
at $\Delta = 384$.

\begin{figure}[ht]
\centering
\includegraphics[width=5.8cm]{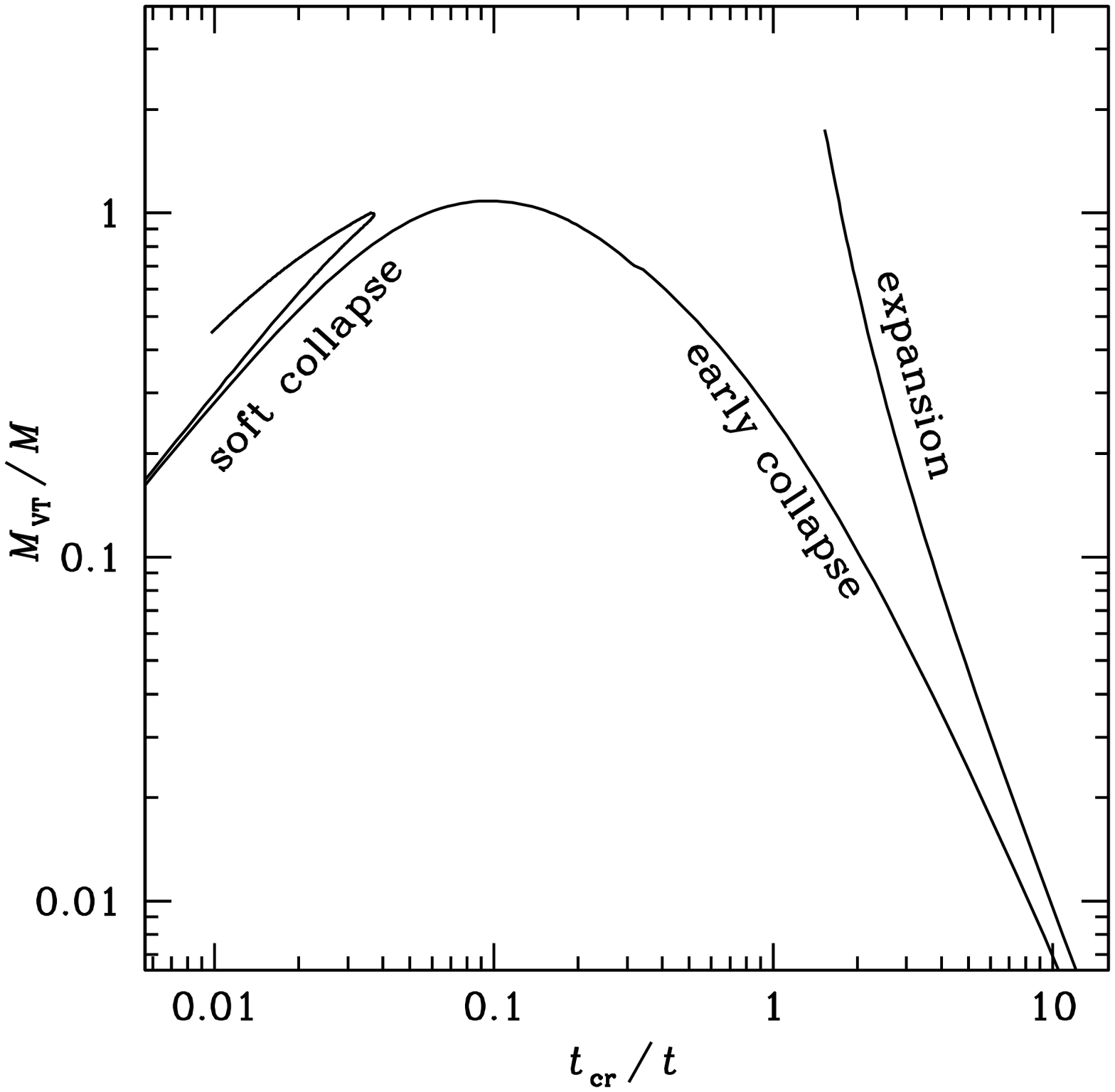}
\includegraphics[width=5.8cm]{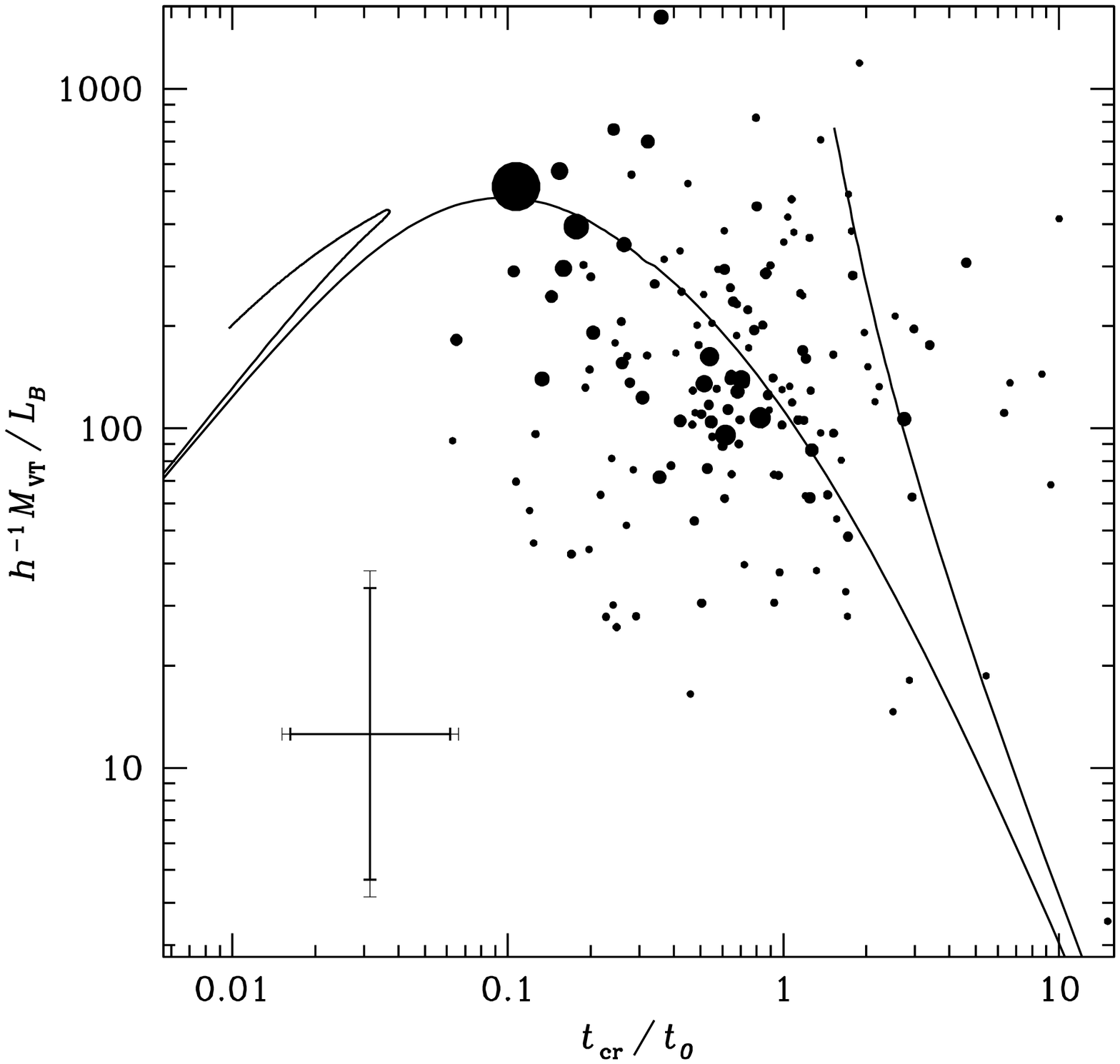}
\caption{Cosmo-dynamical evolution of galaxy systems (adapted from
  \cite{mamon:Mamon93_Aussois_Dyn}). The left panel shows the theoretical evolution
  of an isolated binary with no specific angular momentum and small mass
  ratio. The right panel shows observed groups
  \cite{mamon:FGCP92,mamon:GCF92}, with 
  increasingly larger symbols for richer groups. The \emph{thin} and
  \emph{thick error bars} are for the quartets and typical
  \cite{mamon:FGCP92,mamon:GCF92} groups,
  respectively.
The \emph{largest symbol} is the Virgo cluster.
  The fundamental track was drawn 
  assuming $M_{\rm true}/L_B = 440\,h$.} 
\label{mamon:ftrack1}
\end{figure}

In the extreme, all galaxies of a selected group could be in the infalling
region. 
The departure of such a group from virialization has an important effect on
the derived virial mass of the system
\cite{mamon:Mamon93_Aussois_Dyn,mamon:Mamon94_Moriond,mamon:Mamon95_Chalonge}. 
Indeed, while it is difficult to compare groups of different scale and
cosmo-dynamical state (expansion, collapse, virialization), I had realized
that the dimensionless crossing time, $R/(\sigma_v t)$, and the dimensionless
mass bias, that is mass from the virial theorem divided by true mass, 
$R \sigma_v^2 / (G M)$, provide a plane in which the evolution of isolated
systems in 
an expanding Universe follow a \emph{fundamental track} (FT), 
which is independent
of their mass
\cite{mamon:Mamon93_Aussois_Dyn,mamon:Mamon94_Moriond,mamon:Mamon95_Chalonge}. 
This track is estimated for the case of a binary system of two extended
subgroups of very different
masses, with no specific angular momentum.
The velocity dispersion of a system can be expressed as $\sigma_v^2 = 
({\dot R})^2 + \left (\sigma_v\right)_{\rm proper}^2$. 
The FT, shown in the left panel of Figure~\ref{mamon:ftrack1}, was
derived assuming that the velocity dispersion of the expanding/collapsing
group is dominated by the $\dot R$ term, with the proper velocity dispersion
dominating after the full collapse.

The left panel of 
Figure~\ref{mamon:ftrack1} shows that the true mass is usually severely
underestimated by application of the virial theorem.
During the early stages of collapse, this occurs because the velocity
dispersion of the system is still low (as $\dot R$ is small).
Near full collapse, the galaxies decouple from their dark matter halos:
while the virial theorem measures the mass at the radius of the group where
lie the galaxies, there is dark matter beyond that radius.
The FT diagram (left panel of Fig.~\ref{mamon:ftrack1}) also
indicates that while the crossing time is to first approximation a good
estimator of the cosmo-dynamical evolution of a group, it suffers from
degeneracies between the expansion and early collapse phases, and also
between the full collapse and rebound phases.
These degeneracies can be lifted in part by combining crossing times with
the virial to true mass ratio.
Note that the precise evolution of a system after full collapse is not well
known and probably varies from group to group, depending on the specific
angular momentum of the binary.

The right panel of Figure~\ref{mamon:ftrack1} shows how observed groups relate to
this FT. Since we do not know the true masses of groups, we
assume, as a first order approximation that the optical luminosity of a group
is proportional to its true mass. The true mass-to-light ratio is a free
parameter and is fitted with the highest multiplicity groups, yielding $M_{\rm
  true}/L_B = 440\,h$ (where $h = H_0 / \left [100 \,\rm km \,s^{-1} \,
  Mpc^{-1} \right ]$), while the median $M_{\rm VT}/L_B$ is 4 times smaller
(because it is dominated by low multiplicity groups near turnaround).

Interestingly, \emph{most groups lie near the fundamental track}.
The large scatter for the low multiplicity groups is partly due to larger
errors in 
estimating the plotted quantities given small numbers of galaxies, but the
errors are not sufficient in explaining the large number of low-multiplicity
groups well below the FT, unless these groups, either have
lower $M/L_B < 440\,h$ (because of an intrinsic $M/L$ increase with group
luminosity, or because the galaxies in 
these particular groups are undergoing bursts of star
formation), or are caused by chance alignments of galaxies along the
line-of-sight (for which the group radius is underestimated, while the
velocity dispersion is not).

\begin{figure}[ht]
\centering
\includegraphics[width=7cm]{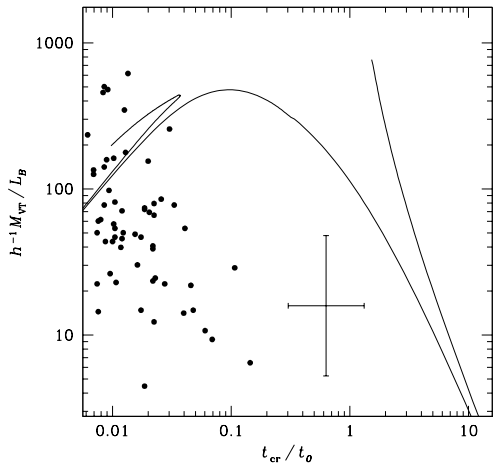}
\caption{Same as right panel of Figure~\ref{mamon:ftrack1}, with the same $M_{\rm
    true}/L_B = 440\,h$,  but for 
Hickson compact groups \cite{mamon:Hickson82,mamon:HMdOHP92}. Velocity dispersion increases
from lower right to upper left.}
\label{mamon:ftrackCG}
\end{figure}
Figure~\ref{mamon:ftrackCG} displays the positions of Hickson's compact groups
\cite{mamon:Hickson82,mamon:HMdOHP92} relative to the FT.
The high velocity dispersion compact groups agree with the predicted
FT for $M_{\rm
  ture}/L_B = 440\,h$, while the low velocity dispersion compact groups do not.
The strong offset of the low $\sigma_v$ compact groups relative to the FT can
be explained 
either by bursts of star formation (making $L_B$ a poor estimator of the
true mass) or by chance alignments of galaxies along the line of sight within
large collapsing groups \cite{mamon:Mamon93_Aussois_Dyn} (for which the compact
group radius is a severe underestimate of the parent group radius).


\section{The evolution of galaxies in groups}
\subsection{The many physical processes at work}

Dense environments are expected to alter the properties of galaxies.
A variety of physical mechanisms are at work in groups and clusters:
basically gravitational and hydrodynamical. We highlight below those
processes that alter the morphologies, gas content (and briefly star
formation efficiency) of galaxies in groups. 

Major galaxy mergers drastically alter the galaxy morphologies, destroying
the disks of spirals galaxies and producing a merger remnant that resembles
an elliptical galaxy, not only in its surface brightness profile
\cite{mamon:Barnes88}, but in the details of its internal kinematics
\cite{mamon:Dekel+05}.
Minor mergers are expected to build up the bulges of spirals at the expense
of their disks \cite{mamon:BJC05}.
But repeated minor mergers can have
the effect of a single major merger \cite{mamon:BJC05}.
Similarly, repeated minor
encounters (merging or not) will ``harass'' a galaxy enough to substantially
transform it \cite{mamon:Moore+96}.
Rapid non-merging encounters also play an important role: the tidal force at
closest approach (pericenter) produces a quadrupolar perturbation on the
shape of the galaxy, and in extreme cases does the same on its giant
molecular clouds. The latter tidal force thus accelerates star formation,
while the former can help gas fall into the galaxy core, thus driving an
episode of active galactic nucleus (AGN) \cite{mamon:BVVS86}.
In addition to these collisional tides, the galaxy is affected by the mean
tidal field of its cluster
\cite{mamon:Merritt84} or dense group \cite{mamon:Mamon87}, where this global
tide is more efficient at altering galaxies than the tides from galaxy
encounters \cite{mamon:Mamon87}.
The mean tidal field of the group or 
cluster does not simply limit the galaxy extent
and mass, but by pumping energy into the reservoir of outer gas, 
it suppresses gas infall onto galaxy disks, thus severely reducing long-term
star formation 
\cite{mamon:LTC80}, a process often called disk strangulation.

Gas can be removed from its parent galaxy through ram pressure
($P~\rho_e\,v^2$, where $\rho_e$ is the gas density of the environment). This
idea was  
first applied to colliding spiral galaxies \cite{mamon:SB51}, and later it
became clear that the hot intracluster gas provides an important ram pressure
\cite{mamon:GG72}. 
After losing its gas, a spiral galaxy would see its disk fade and resemble a
lenticular (S0) galaxy.
It is not yet clear that S0s are indeed produced by ram pressure gas
stripping, or are simply systems with large bulge to disk ratio.
Ram pressure stripping produces the correct local and global trends in
clusters \cite{mamon:SS92}, but 
cannot explain the presence of S0s in groups, where the lower velocities are
unable to produce sufficiently high ram pressure 
\cite{mamon:Dressler86,mamon:HP03_mdr}.
On the other hand, if the large bulges of S0s are built up by minor mergers,
the fraction of S0s is clusters are found to be lower in semi-analytical
models of galaxy formation \cite{mamon:SWTK01} than observed. However, the
inclusion of ram pressure stripping (in very simplistic schemes) does not
increase enough the S0 fraction in semi-analytical models
\cite{mamon:ON03,mamon:Lanzoni+05}. 

The morphological segregation of galaxies in groups and clusters obviously
leads to color segregation, where the redder galaxies lie in the dense
central regions, but interestingly, the galaxy colors are redder for given
morphological type in clusters than in the field \cite{mamon:Balogh+98},
which can 
be simply explained by the tidal dissipation of the gas reservoirs of spirals
\cite{mamon:BNM00}.

Most of the processes listed here work best in dense environments. On the
contrary, for 
very loose groups such as the Local Group, which are still in the collapse
phase, major interactions have not yet altered the more massive
galaxies. Only the low mass satellites of massive galaxies are affected by the
presence of their more massive companion.


\subsection{How frequently do group galaxies merge?}

The merger rate per unit volume 
of equal mass galaxies in a group is obtained by integrating
over a merger cross-section:
\begin{equation}
k(m) \equiv k(m,m) = \left\langle v S(v) \right \rangle
= \int_0^\infty v S(v) f(v)\,dv \ ,
\end{equation}
where $S(v) = \pi\,p_{\rm crit}^2(v)$ is the relative-velocity dependent
merger cross-section and $f(v)$ is the distribution of relative velocities. 
For a linear $p_{\rm crit}(v)$
\cite{mamon:RN79}, one finds  
\cite{mamon:Mamon92}
\begin{equation}
k = b\,{r_h^2\,\sigma_g^4\over\sigma_e^3} 
\ ,
\label{mamon:k1}
\end{equation}
where $r_h$ and $\sigma_g$ are the half-mass radius and (1D) velocity
dispersion  of the galaxies, $\sigma_e$ is the velocity dispersion of the
environment, and with now old cross-sections scalings
\cite{mamon:RN79,mamon:AF80} one 
has $b \simeq 200$. 
With 
$\sigma_g^2 \simeq {0.4/ 3}\,{G m / r_h}$ \cite{mamon:Spitzer69},
one then gets \cite{mamon:Mamon92}
\begin{equation}
k(m) = a\,{G^2\,m^2\over \sigma_e^3} \ ,
\label{mamon:k1m}
\end{equation}
where $a \simeq 3.5$.
The merger rate has also been derived directly
\cite{mamon:MH97} 
from fairly realistic
$N$-body simulations of the mergers
of equal mass Hernquist
\cite{mamon:Hernquist90} models, and this rate is essentially
identical to my analytical prediction \cite{mamon:Mamon92}, as shown in a previous
review of mine \cite{mamon:Mamon00_IAP}, despite my 
neglect of gravitational focussing and my use of merger
cross-sections based upon very primitive $N$-body simulations
\cite{mamon:RN79,mamon:AF80}. 

In my previous review \cite{mamon:Mamon00_IAP}, I've adapted
equation~(\ref{mamon:k1}) to estimate the
merger rates for unequal mass galaxies, and integrating over the mass
function of field galaxies I derived a merger rate as a function of galaxy mass:
\begin{equation}
{dN_m\over dt} = \hbox{cst}\,n_* {G^2 m_*^2 \over \sigma_e^3}\,{\cal R} \left
({m\over m_*} \right ) \ ,
\label{mamon:direct}
\end{equation}
where $m_*$ and $n_*$ are the break in the galaxy mass function and a
fiducial number density, while ${\cal R}(m)$ is a dimensionless function of
mass which, for major mergers, increases with mass, reaches a maximum near
$m=m_*$ and decreases sharply for $m>m_*$.
The normalization is such that 
$m_*$ galaxies in typical $\sigma_v = 300 \, \rm km \, s^{-1}$
groups should suffer a few major mergers per Hubble time
\cite{mamon:Mamon00_IAP}.
I also
estimated merger rates as a function of position in the group or cluster and
found that the direct merger rate outside of the the central galaxy is
maximum at $\simeq 0.1\,R_{200}$ \cite{mamon:Mamon00_IAP}.
An increasing number of observational constraints on the galaxy merger rate
and its evolution are now arising (see Conselice in these proceedings and
\cite{mamon:Conselice06}) and need to be compared with the predictions given
here. 

\subsection{Do we understand the morphology-density relation in groups and
  clusters?}  

The predictions above on the rates of direct mergers versus environment can
be compared to the global morphological mix and its local variations (the
morphology-density relation, hereafter MDR) observed in groups and 
clusters.
While an early analysis \cite{mamon:PG84} concludes to a universal MDR 
as a function of local galaxy density, a closer look reveals various
problems with groups of galaxies: Hickson's \cite{mamon:Hickson82} compact
groups, assumed as dense in 3D as they appear in 2D, have much higher spiral
fractions than expected from the universal MDR 
\cite{mamon:Mamon86}.  Conversely, these compact groups are spiral-poor
relative 
to other systems of the same velocity dispersion \cite{mamon:HR88}, but with a
very strong inverse correlation of spiral fraction and velocity dispersion
\cite{mamon:HKH88}, 
also seen in general SDSS groups \cite{mamon:WvdBYM06}.

Recently, Helsdon \& Ponman \cite{mamon:HP03_mdr} compared
the local morphological mix of their sample of X-ray emitting groups
to that of clusters \cite{mamon:Dressler+97}.
After statistical correction for projections, X-ray emitting groups appear to
be spiral-poor relative to clusters of the same local 3D density, by an
amount just as expected from the $\sigma_e^{-3}$ scaling of
equations~(\ref{mamon:k1}) and (\ref{mamon:k1m})
\cite{mamon:HP03_mdr}.
In other words, the true groups, as selected by their X-ray emission, have
less spirals than clusters because their lower velocity dispersions lead to
slow enough encounters to lead to major mergers that transform spirals into
ellipticals. 

This simple understanding of the MDR of groups vs
clusters has 3 caveats:
\begin{enumerate}
\itemsep 2pt
\item Major mergers are probably more often the result of orbital decay by 
 dynamical friction
  than of direct slow collisions. The orbital decay time
 scales as the dynamical friction (df) time \cite{mamon:Mamon95_Chalonge}, which  
scales as
\begin{equation}
\left ({dN_m\over dt}\right )_{\rm df} \propto
\tau_{\rm df}^{-1} \propto {G^2\,m\,\rho\over \sigma_e^3}\,\ln
\left({M(R)\over m}\right) \ .
\label{mamon:fric}
\end{equation}
Comparing equations~(\ref{mamon:direct}) and (\ref{mamon:fric}), one finds
that the rate of direct to frictional mergers scales as $\ln (M(R)/m)/f$,
where $f$ is the mass fraction of the group/cluster in galaxies.
This small (logarithmic) dependence on the mass of the environment means that
\emph{to first order, the ratio of frictional to direct mergers is
  independent of the environment}, with direct mergers slightly more (less)
  important in groups (clusters).
\item When two groups or clusters merge, violent relaxation might
  cause their galaxy populations to mix sufficiently to erase their
  MDRs. However, 
$N$-body simulations indicate that when two systems merge,
  there is a strong 
  correlation between initial and final binding energies
  \cite{mamon:Barnes92}, so that the most 
  bound galaxies in the initial groups or clusters, mostly ellipticals, will
  end up as the most bound galaxies, thus preserving the MDR.
\item Our analysis pertains to instantaneous merger rates. Since we live in a
  hierarchical Universe where the rich clusters today were built of smaller
  groups, we still need to
  check that the ratio of instantaneous merger rates in groups and clusters
  is equal to the ratio of the time-averaged merger rates in groups and
  clusters. But since the merger rates scales as $\sigma_e^{-3}$, hence as
  one over the 
  mass of the environment,
the ratio at any time between the merger rates of the main progenitors of 
present-day groups
  and clusters will scale as $M_{\rm group}(t)/M_{\rm cluster}(t)$.
Since massive clusters are rare objects, they must form recently,
  because otherwise they would be even more extreme objects in the
  past. However, this effect is not too severe: from cosmological $N$-body
  simulations and analytical predictions of the mass assembly
  history of cosmic systems  \cite{mamon:vandenBosch02}, I infer that the
  elapsed time for the doubling in mass of (the main progenitors of) 
present-day groups ($z=0.9$) is only 1/3 greater than that of present day
  clusters ($z=0.4$). Therefore, the average mass ratio of the group to
  cluster progenitors 
  decreases only slightly with time, so that \emph{to first order the time
  averaged 
  ratio of merger rates of group and cluster progenitors is close to the
 ratio of present-day merger rates}.

\end{enumerate}
Therefore, none of the caveats is serious, and it does appears that
\emph{the increased rate of galaxy mergers indeed explains 
the larger local fractions of ellipticals in X-ray selected groups relative
to clusters}.

Since compact groups in general are spiral-rich \cite{mamon:Mamon86}, while
X-ray 
selected groups are spiral-poor \cite{mamon:HP03_mdr}, then \emph{the compact
  groups that are 
not in the GEMS sample must be nearly devoid of early-type galaxies},
i.e. the spiral fractions of these groups should be as high or higher 
than in the
average field. The non-X-ray emitting compact groups are typically low
velocity dispersion groups, which as we saw, do not fit the FT, so it is
tempting to conclude that \emph{low velocity dispersion 
  compact groups are mostly caused by chance projections}.

\subsection{How far out should galaxies feel the group environment?}

The decreased star formation rate for group/cluster galaxies of given
morphological type is visible out to $2\,r_{200}$ \cite{mamon:Balogh+98},
i.e. $\approx 2.6\,r_{100}$. 
Regardless of 
whether the mass assembly of groups and clusters is viewed in a
monolithic spherical infall context or in a hierarchical merging one,
the effects of the group/cluster on the galaxies should
thus be seen out to the maximum \emph{rebound} or \emph{backsplash} radius.
If this radius is assimilated to the radius of mixing in the spherical infall
model, one finds $r_{\rm reb} \approx r_{\rm vir}$ \cite{mamon:MSSS04}.
If one assumes values for the ratio of the (lagrangian) rebound to turnaround
radius (e.g. $1/2$), and for the corresponding ratio of rebound to turnaround
times (e.g. $3/2$), one can solve in the context of a flat Universe with a
cosmological constant for the rebound radius, and we find $r_{\rm reb} /
r_{100} \approx 1$ with a maximum of 2.5 for the most favorable case (rebound
radius equals turnaround
radius and rebound time equals twice the turnaround time) \cite{mamon:MSSS04}.
Deriving reasonable ratios from particle trajectories of an $\Lambda$CDM  
simulation
\cite{mamon:FM01}, we obtain $r_{\rm reb} / r_{100}$ ranging from 0.55 to
1.25 \cite{mamon:MSSS04}.
Finally we considered the 
final output ($z=0$) of a GALICS galaxy formation simulation
\cite{mamon:Hatton+03} built on top of a dark matter only $\Lambda$CDM cosmological
simulation, where galaxies penetrating groups and clusters are selected by
empty halos (a feature of GALICS is that when halos merge, galaxies remain
with the more massive halo). We then found that galaxies (empty halos) travel
out to $1.7\,r_{100}$ \cite{mamon:MSSS04}.
These numbers were confirmed through an analysis of the particle trajectories
in $\Lambda$CDM halos, which shows that particles that penetrate deep into
their 
halo travel out to $2\,r_{100}$  \cite{mamon:GKG05} $\simeq 2.6 \,r_{200}$.
These maximum rebound radii are thus somewhat larger (by 30\%)
than the observed maximum radius for decreased star formation, 
but consistent this value.


\section{The internal kinematics of groups}
Clusters of galaxies are beginning to reveal their mass profiles, through
analyses of their internal kinematics, X-ray gas or lensing properties.
Kinematical analyses are often based upon the Jeans equation of hydrostatic
equilibrium, relating the divergence of the anisotropic dynamical pressure
tensor with the tracer density times the gradient of the gravitational
potential. In spherical symmetry this becomes 
\begin{equation}
{\D \left (\nu \sigma_r^2\right) \over \D r} + 2\,\beta\,{\nu \sigma_r^2 \over
  r} = - \nu {G M(r) \over r^2} \ ,
\end{equation}
where $\nu(r)$, $M(r)$ and $\beta(r)$ are the radial profiles of tracer
number density, total mass, and velocity anisotropy $\beta = 1 -
\sigma_\theta^2/\sigma_r^2$ (where $\beta=1, 0$, and $\to-\infty$ corresponds
to radial, isotropic and
circular orbits, respectively).
Since the data are seen in projection, one has to couple the Jeans equation
with the anisotropic projection equation
\cite{mamon:BM82}
\begin{equation}
\Sigma(R)\,\sigma_{\rm los}^2 (R) = 
2\,G\,\int_R^\infty \left (1 - \beta\,{R^2\over r^2} \right )
\nu \sigma_r^2 \,{r\,\D r\over \sqrt{r^2-R^2}} \ ,
\label{anisproj}
\end{equation}
where $\Sigma(R)$ is the surface number density as a function of projected
radius $R$, and 
$\sigma_{\rm los}(R)$ is the line of sight velocity dispersion profile
(hereafter VDP).

For example, assuming isotropic to moderately radial
orbits, the mass profile of clusters is consistent with
being proportional to the galaxy number density profile \cite{mamon:CYE97}, 
which itself was
consistent with the cuspy NFW \cite{mamon:NFW96} profile.
Recently, the mass-anisotropy degeneracy, inherent in the Jeans equation, was
recently (partially) 
lifted: the early type
galaxies in clusters follow isotropic orbits \cite{mamon:LM03,mamon:KBM04},
galaxy velocities are isotropic in groups and somewhat radial in clusters
\cite{mamon:MG04}. 

Andrea Biviano and I are currently
performing a similar analysis for groups of galaxies, as we wish to
check if various types of groups (hot vs. cold, etc.) display similar mass
profiles and concentrations. For this, we use the GEMS group sample
\cite{mamon:OP04}, which considers all groups for which there have been X-ray
pointings. In this review, we focus on the groups with diffuse X-ray emission
distinct from diffuse emission around the central bright elliptical.

We selected group member galaxies with NED (which at
the time included the SDSS-DR4 and the 6dFGS-DR2),
searching out to twice (to be conservative) a 1st-order estimate of their
virial radius,
$r_{200}$ (where the mean density is 200 times the critical density of the
Universe) using $r_{200}/\hbox{Mpc} =
\sigma_v/450 
\, \rm km \, s^{-1}$, as appropriate for pure NFW models with $c=8$ and $H_0
= 70 \,\rm km \,s^{-1} \, Mpc^{-1}$ (when $\sigma_v < 300 \, \rm km \,
s^{-1}$, we used $\sigma_v = 300 \, \rm km \, s^{-1}$ to be conservative) and
out to $\pm 5\,\sigma_v$. 
We then computed the velocity dispersion and iterated with a $\pm
3\,\sigma_v$ depth.

\subsection{Which is the best estimator of virial radius?}

Since groups have few galaxies, we need to stack them, normalizing the radii
to their virial radius, $r_{200}$ (as first done by
\cite{mamon:Carlberg+01_groups}), 
and similarly stack their velocities to
$V_{200}$, the circular velocity at $r_{200}$.
We have considered several estimators of the virial radius:
1) the velocity dispersion, assuming an isotropic pure NFW model
with a $\Lambda$CDM concentration parameter, $\sigma$-NFW; 2) the $K$-band
total galaxy 
luminosity, taken from the 2MASS survey, corrected for incompleteness, using
an $M-L_K$ relation
\cite{mamon:LMS03}; 
3) the X-ray emission-weighted temperature, using $M-T$
relations from several authors
\cite{mamon:EMN96,mamon:Sanderson+03,mamon:APP05}. 
For each of our $r_{200}$ estimators, we performed the interloper removal,
first group per group, then on the stacked pseudo-group. We also removed
a few groups contaminated by nearby groups or clusters and thse with $<5$
members.
In the stacking, we have weighted the galaxies inversely proportional to the
number of galaxies in the group out to the virial radius. This weighting
gives equal weight to each group in the stacked pseudo-group.

\begin{figure}[ht]
\centering
\includegraphics[width=10cm]{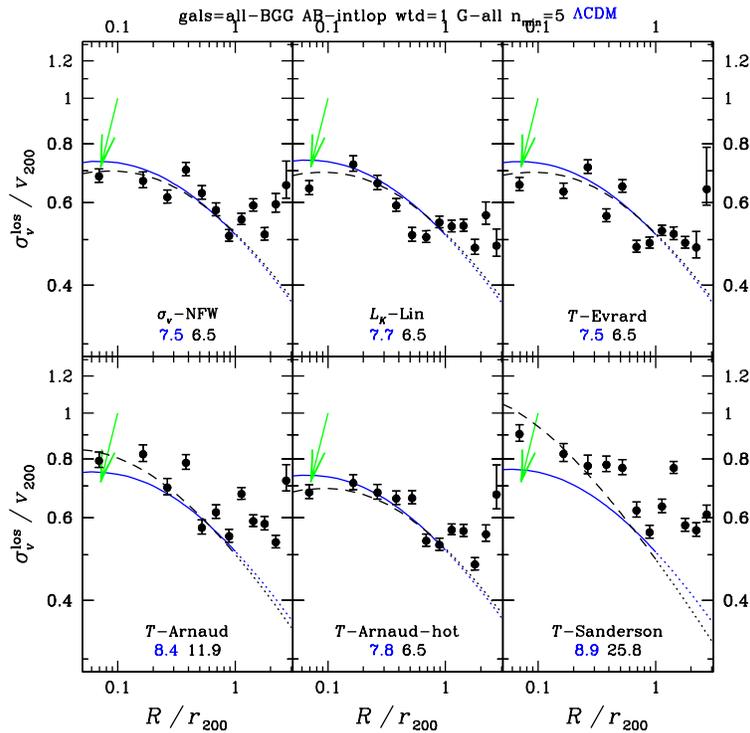}
\caption{Line-of-sight velocity dispersion profiles for GEMS groups with
  diffuse emission, drawn for 6 samples stacked using different methods for
  estimating the virial radius.
The curves are the predictions (eq.~[\ref{mamon:Isiglosgen}]) for isotropic
  systems 
  with NFW profiles: total mass and number number densities following
  $\Lambda$CDM predictions (\emph{solid (blue)}), as well as 
  different NFW fits to
  total mass and number density profiles (\emph{dashed (black)}).
The \emph{dotted curves} show the extrapolations beyond the last fitted radius
  ($R=r_{200}$).
The \emph{arrows} indicate the effect of underestimating the virial radius by
  a factor 1.4.
The \emph{labels} at the bottom indicate the method to estimate $r_{200}$,
  where 
  {\sf $T$-Arnaud-hot} refers to the fit to the hotter clusters of the Arnaud
  et 
  al. sample.
The \emph{numbers under the labels} indicate the concentration parameters for
  the $\Lambda$CDM model and the best NFW fit to the total mass density.
}
\label{mamon:siglos6}
\end{figure}

Figure~\ref{mamon:siglos6} compares the measured VDPs with the expectation from
the isotropic Jeans equation \cite{mamon:PS97}:
\begin{equation}
\Sigma(R)\,\sigma_{\rm los}^2 (R) = 2\,G\,R\,
\int_R^\infty 
\sqrt{1-{R^2 \over r^2}}
\,\nu(r)\,M(r)\,{\D r\over r} 
\ .
\label{mamon:Isiglosgen}
\end{equation}
We are also investigating other simple 
anisotropy profiles for which the term in the
square root is replaced by other kernels given by \cite{mamon:ML05b}, and
alternatively, we are directly estimating the mass profile for given
anisotropy profiles using the recently discovered mass inversion technique
\cite{mamon:MB06}. 

The Sanderson et al. 
\cite{mamon:Sanderson+03} $M-T$ relation,
which predicts lower mass groups than when extrapolated from cluster $M-T$
relations, 
 appears inconsistent with
isotropic NFW models, as the virial radius and velocity dispersion appear to be
overestimated by a factor 1.3, i.e. the mass is a factor of 2 higher than
predicted from their $M-T$ relation.
A maximum likelihood analysis at $\beta=\rm cst$
suggests that the Sanderson et al.
$M-T$ relation can only be reconciled with a highly concentrated ($c\approx
100$) and tangentially anisotropic ($\beta\approx -2$) system.
On the other hand, the virial radii built upon $M-T$ relations derived for hot
clusters \cite{mamon:APP05} or cosmological simulations
\cite{mamon:EMN96} produce velocity dispersion profiles that
are consistent with isotropic NFW models.
Therefore, \emph{the $M-T$ relation of groups of galaxies
  appears to lie in the extrapolation of that of clusters}, with no break at
$\approx 10^{13} M_\odot$.

\subsection{Constraints on group mass profiles}

\begin{figure}[ht]
\centering
\includegraphics[width=10cm]{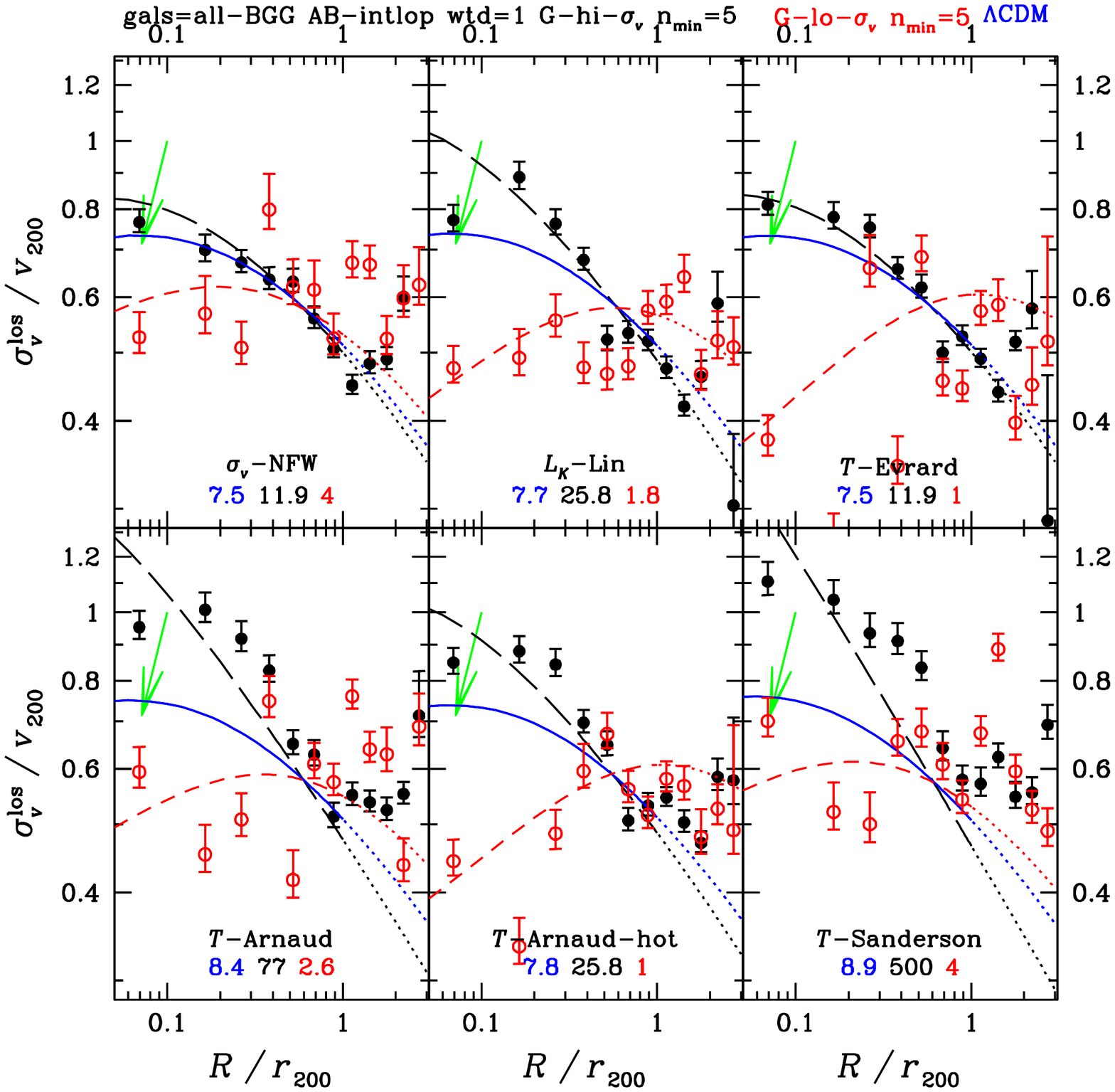}
\caption{Same as
  Figure~\ref{mamon:siglos6}, with \emph{long-dashed (black)} and 
  \emph{short-dashed (red) curves} displaying the fits for the dynamical hot
($\sigma_v > 300 \, \rm km \, s^{-1}$) and cold ($\sigma_v \leq 300 \, \rm km
\, s^{-1}$) groups.
The \emph{three numbers under the labels} give the NFW concentration parameters
  for the $\Lambda$CDM model as well as the best fit to the total mass
  density profiles of the dynamical hot and cold groups, respectively.
}
\label{mamon:siglos6sigv}
\end{figure}

We have subdivided the G groups into subclasses according to their velocity
dispersion, $\sigma_v$, temperature,  $T$, ratio of galaxy orbital to gas
thermal energies $\beta_{\rm spec} =
\sigma_v^2/(kT/\mu m_p)$, X-ray luminosity, $L_X$ and $K$-band luminosity,
$L_K$. 
Figure~\ref{mamon:siglos6sigv} shows plots analogous to those in
Figure~\ref{mamon:siglos6}, but when the sample is subdivided into dynamically hot
($\sigma_v > 300 \, \rm km \, s^{-1}$) and cold ($\sigma_v \leq 300 \, \rm km
\, s^{-1}$) groups.
Regardless of the method used to estimate $r_{200}$, 
\emph{the dynamically hot (cold) groups
are always more (less) concentrated in total mass
than the $\Lambda$CDM prediction for
isotropic orbits}.
While clusters of galaxies display mass concentrations that are consistent
with the $\Lambda$CDM predictions (\cite{mamon:APP05} using X-rays and
\cite{mamon:BG03,mamon:LM03,mamon:KBM04,mamon:Lokas+06} using internal
kinematics), the richer 
(high velocity dispersion) groups appear to be more concentrated than
$\Lambda$CDM predictions, 
while the poorer (low velocity dispersion) groups appear to be less
concentrated than $\Lambda$CDM predictions.
The high concentrations of the hot groups are caused in part by fitting the
total mass density profile, while the more concentrated baryonic (S\'ersic)
component gives the illusion of high total mass concentrations
\cite{mamon:ML05a}. 
Note that the galaxy distribution is even more concentrated, leading to
rising mass over number ratio, consistent with what was derived on an
optically selected sample \cite{mamon:Carlberg+01_groups}.
Alternatively, hot (cold) groups could have $\Lambda$CDM mass profiles,
with  radial
(tangential) galaxy orbits.
Finally, the strong scatter in the VDPs of the dynamically cold groups
suggests that \emph{dynamically cold groups are contaminated by unreal groups}.

There are also strong differences between groups with high and low
$\beta_{\rm spec}$, and X-ray luminosity, but smaller differences when
subdividing groups into classes of high and low X-ray temperature or $K$-band
luminosity. 

\subsection{The fundamental track of groups selected to the virial radius}

\begin{figure}[ht]
\centering
\includegraphics[width=9.2cm]{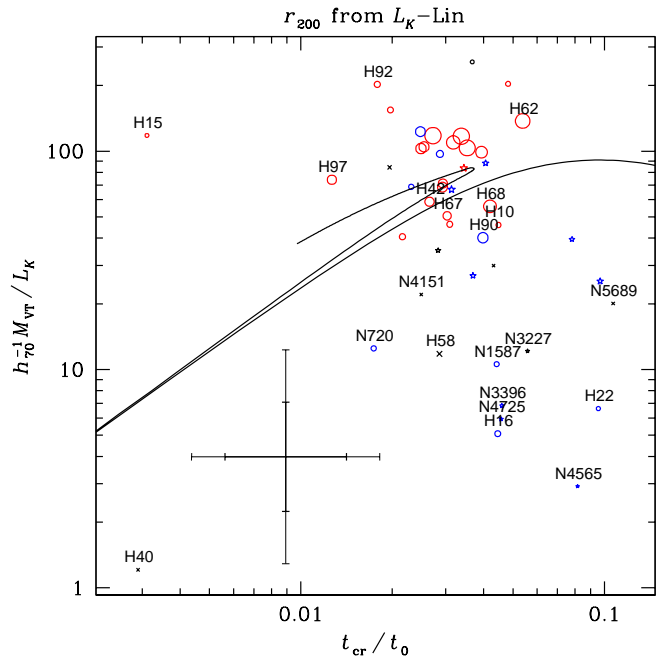}
\caption{Same as right panel of Figure~\ref{mamon:ftrack1}, for GEMS groups
  selected out to 
  the virial radius (defined with the $K$-band luminosity \cite{mamon:LMS03}). 
The \emph{open symbols}, \emph{stars} and \emph{crosses} indicate the GEMS G
  (extended X-ray emission distinct from central galaxy emission), H (extended
  X-ray emission indistinguishable from that of central galaxy) and U
  (undetected extended X-ray emission) groups, respectively. The \emph{red},
  \emph{blue} and \emph{black symbols} are for high ($>0.5$), low ($\leq
  0.5$) and 
  undefined (unavailable temperature)
$\beta_{\rm spec}$, respectively.
The FT was placed assuming the universal value $M/L_K =
  84\,h_{70}$ as derived from the $K$-band luminosity density of the 6dFGS
  galaxies \cite{mamon:JPCS06}.
The \emph{symbol sizes} are proportional to the square root of the number of
  galaxies within $r_{200}$.
Outliers to the FT and Hickson compact groups are highlighted.
}
\label{mamon:ftrackGEMS}
\end{figure}

Figure~\ref{mamon:ftrackGEMS} shows the FT of 
GEMS groups, which were selected out to the
virial radius, here defined with the $L_K$ luminosity.
The rich groups cluster near the position expected for virialized systems
(the kink of the FT at $t_{\rm cr} = 0.035\,t_0$).
Interestingly, the groups that lie off the FT (at lower $M_{\rm
  VT}/L_K$) have typically low $\beta_{\rm spec}$, but also longer crossing
times. This suggests that these 
low velocity dispersion groups are not simply low mass-to-light ratio
systems, but that there is an intrinsic property that takes them off the FT.
The simplest explanation is that \emph{low 
velocity dispersion groups (compact or not)
are chance alignments of galaxies
along the line of sight, i.e. prolate groups in real space}.

\section{Concluding thoughts}
Our understanding of  the evolution
of groups and of galaxies therein is making rapid progress thanks to 
the advent of 1) large galaxy surveys such as the 2dFGRS and SDSS,
2) multi-wavelength observations of groups,  and 3)
high resolution cosmological $N$ body simulations.
Many of the 
results presented here need to
be confirmed with these large data sets and simulation outputs.

I am grateful to my collaborator, Andrea Biviano,
for
allowing me to mention our work in progress. I also thank Ivo Saviane, Valentin
Ivanov and Jordanka Borissova for organizing a very exciting and high-level
meeting and for being extremely patient with the manuscript.
%

\begin{thebibliography}{99}

\bibitem{mamon:OP04}
J.P.F. {Osmond}, T.J. {Ponman}:
MNRAS \textbf{350}, 1511 (2004)

\bibitem{mamon:SMCB93}
R.~{Schaeffer}, S.~{Maurogordato}, A.~{Cappi}, F.~{Bernardeau}:
MNRAS \textbf{263}, L21 (1993)

\bibitem{mamon:MFW93}
B.~{Moore}, C.S. {Frenk}, S.D.M. {White}:
MNRAS \textbf{261}, 827 (1993)

\bibitem{mamon:DKCW99}
A.~{Diaferio}, G.~{Kauffmann}, J.M. {Colberg}, S.D.M. {White}:
MNRAS \textbf{307}, 537 (1999 )

\bibitem{mamon:Spergel+06}
D.N. {Spergel}, R.~{Bean}, O.~{Dor\'e}, et al.:
ApJ submitted, arXiv:astro-ph/0603449 (2006)

\bibitem{mamon:KS96}
T.~{Kitayama}, Y.~{Suto}:
ApJ \textbf{469}, 480 (1996)

\bibitem{mamon:GH83}
M.J. {Geller}, J.P. {Huchra}:
ApJS \textbf{52}, 61 (1983)

\bibitem{mamon:RGH89}
M.~{Ramella}, M.J. {Geller}, J.P. {Huchra}:
ApJ \textbf{344}, 57 (1989)

\bibitem{mamon:MZ02}
M.~{Merch{\'a}n}, A.~{Zandivarez}:
MNRAS \textbf{335}, 216 (2002)

\bibitem{mamon:MZ05}
M.E. {Merch{\'a}n}, A.~{Zandivarez}:
ApJ \textbf{630}, 759 (2005)

\bibitem{mamon:Mamon93_Aussois_Dyn}
G.A. {Mamon}:
Dynamical theory of groups and clusters of galaxies.
In: \textit{Gravitational Dynamics and the N-Body Problem}, 
ed by F.~{Combes}, E.~{Athanassoula} (Obs. de Paris, Paris 1993) 
pp 188--203,
arXiv:astro-ph/9308032

\bibitem{mamon:Mamon94_Moriond}
G.A. {Mamon}:
The galaxy group/cosmology connections. 
In: Moriond Astrophysics Mtg number~14,
\textit{Clusters of Galaxies}, 
ed by F.~{Durret}, A.~{Mazure}, S.D.M. {White}, J.~{Tr\^anh Thanh 
  V\^an}  
(Fronti\`eres, Gif-sur-Yvette 1994)
pp 291--296, 
arXiv:astro-ph/9406043

\bibitem{mamon:Mamon95_Chalonge}
G.A. {Mamon}:
The dynamics of groups and clusters of galaxies and links to cosmology.
In: 3rd Paris cosmology colloq,
ed by H. de Vega, N. S\'anchez
(World Scientific, Singapore 1995) 
pp 95--119,
arXiv:astro-ph/9511101

\bibitem{mamon:FGCP92}
P.~{Fouqu\'e}, E.~{Gourgoulhon}, P.~{Chamaraux}, G.~{Paturel}:
A\&AS \textbf{93}, 211 (1992)

\bibitem{mamon:GCF92}
E.~{Gourgoulhon}, P.~{Chamaraux}, P.~{Fouqu\'e}:
A\&A \textbf{255}, 69 (1992)

\bibitem{mamon:Hickson82}
P.~{Hickson}:
ApJ \textbf{255}, 382 (1982)

\bibitem{mamon:HMdOHP92}
P.~{Hickson}, C.~{Mendes de Oliveira}, J.P. {Huchra}, G.G. {Palumbo}:
ApJ \textbf{399}, 353 (1992)

\bibitem{mamon:Barnes88}
J.E. {Barnes}:
ApJ \textbf{331}, 699 (1988)

\bibitem{mamon:Dekel+05}
A.~{Dekel}, F.~{Stoehr}, G.A. {Mamon}, T.J. {Cox}, G.S. {Novak}, J.R.
  {Primack}:
Nature \textbf{437}, 707 (2005)

\bibitem{mamon:BJC05}
F.~{Bournaud}, C.J. {Jog}, F.~{Combes}:
A\&A \textbf{437}, 69 (2005)

\bibitem{mamon:Moore+96}
B.~{Moore}, N.~{Katz}, G.~{Lake}, A.~{Dressler}, A.~{Oemler}, Jr.:
Nature \textbf{379}, 613 (1996)

\bibitem{mamon:BVVS86}
G.G. {Byrd}, M.J. {Valtonen}, L.~{Valtaoja}, B.~{Sundelius}:
A\&A \textbf{166}, 75 (1986)

\bibitem{mamon:Merritt84}
D.~{Merritt}:
ApJ \textbf{276}, 26 (1984)

\bibitem{mamon:Mamon87}
G.A. {Mamon}:
ApJ \textbf{321}, 622 (1987)

\bibitem{mamon:LTC80}
R.B. {Larson}, B.M. {Tinsley}, C.N. {Caldwell}:
ApJ \textbf{237}, 692 (1980)


\bibitem{mamon:SB51}
L.J. {Spitzer}, W.~{Baade}:
ApJ \textbf{113}, 413 (1951)

\bibitem{mamon:GG72}
J.E. {Gunn}, J.R. {Gott}:
ApJ \textbf{176}, 1 (1972)

\bibitem{mamon:SS92}
J.M. {Solanes}, E.~{Salvador-Sol\'e}:
ApJ \textbf{395}, 91 (1992)

\bibitem{mamon:Dressler86}
A.~{Dressler}:
ApJ \textbf{301}, 35 (1986)

\bibitem{mamon:HP03_mdr}
S.F. {Helsdon}, T.J. {Ponman}:
MNRAS \textbf{339}, L29 (2003)

\bibitem{mamon:SWTK01}
V.~{Springel}, S.D.M. {White}, G.~{Tormen}, G.~{Kauffmann}:
MNRAS \textbf{328}, 726 (2001)

\bibitem{mamon:ON03}
T.~{Okamoto}, M.~{Nagashima}:
ApJ \textbf{587}, 500 (2003)

\bibitem{mamon:Lanzoni+05}
B.~{Lanzoni}, B.~{Guiderdoni}, G.A. {Mamon}, J.~{Devriendt}, S.~{Hatton}:
MNRAS \textbf{361}, 369 (2005)

\bibitem{mamon:Balogh+98}
M.L. {Balogh}, D.~{Schade}, S.L. {Morris}, H.K.C. {Yee}, R.G. {Carlberg}, 
  E.~{Ellingson}:
ApJL \textbf{504}, L75 (1998)

\bibitem{mamon:BNM00}
M.L. {Balogh}, J.F. {Navarro}, S.L. {Morris}:
ApJ \textbf{540}, 113 (2000)

\bibitem{mamon:RN79}
N.~{Roos}, C.A. {Norman}:
A\&A \textbf{76}, 75 (1979)

\bibitem{mamon:Mamon92}
G.A. {Mamon}:
ApJL \textbf{401}, L3 (1992)

\bibitem{mamon:AF80}
S.J. {Aarseth}, S.M. {Fall}:
ApJ \textbf{236}, 43 (1980)

\bibitem{mamon:Spitzer69}
L.~{Spitzer}:
ApJL \textbf{158}, L139 (1969)

\bibitem{mamon:MH97}
J.~{Makino}, P.~{Hut}:
ApJ \textbf{481}, 83 (1997)

\bibitem{mamon:Hernquist90}
L.~{Hernquist}:
ApJ \textbf{356}, 359 (1990)

\bibitem{mamon:Mamon00_IAP}
G.A. {Mamon}:
Theory of galaxy dynamics in clusters and groups.
In: 15th IAP Astrophys. Mtg., 
\textit{Dynamics of Galaxies: from the Early Universe to the Present},
vol 197,
ed by F.~{Combes}, G.A. {Mamon}, V.~{Charmandaris}
(ASP, San Francisco 2000)
pp 377--387,
arXiv:astro-ph/9911333

\bibitem{mamon:Conselice06}
C.J. {Conselice}:
ApJ \textbf{638}, 686 (2006)

\bibitem{mamon:PG84}
M.~{Postman}, M.J. {Geller}:
ApJ \textbf{281}, 95 (1984)

\bibitem{mamon:Mamon86}
G.A. {Mamon}:
ApJ \textbf{307}, 426 (1986)

\bibitem{mamon:HR88}
P.~{Hickson}, H.J. {Rood}:
ApJL \textbf{331}, L69 (1988)

\bibitem{mamon:HKH88}
P.~{Hickson}, E.~{Kindl}, J.P. {Huchra}:
ApJ \textbf{331}, 64 (1988)

\bibitem{mamon:WvdBYM06}
S.M. {Weinmann}, F.C. {van den Bosch}, X.~{Yang}, H.J. {Mo}:
MNRAS \textbf{366}, 2 (2006)

\bibitem{mamon:Dressler+97}
A.~{Dressler}, A.J. {Oemler}, W.J. {Couch}, et al.:
ApJ \textbf{490}, 577 (1997)

\bibitem{mamon:Barnes92}
J.E. {Barnes}:
ApJ \textbf{393}, 484 (1992)

\bibitem{mamon:vandenBosch02}
F.C. {van den Bosch}:
MNRAS \textbf{331}, 98 (2002)

\bibitem{mamon:MSSS04}
G.A. {Mamon}, T.~{Sanchis}, E.~{Salvador-Sol{\' e}}, J.M. {Solanes}:
A\&A \textbf{414}, 445 (2004)

\bibitem{mamon:FM01}
T.~{Fukushige}, J.~{Makino}:
ApJ \textbf{557}, 533 (2001)

\bibitem{mamon:Hatton+03}
S.~{Hatton}, J.~{Devriendt}, S.~{Ninin}, F.R. {Bouchet}, B.~{Guiderdoni}, 
  D.~{Vibert}:
MNRAS \textbf{343}, 75 (2003)

\bibitem{mamon:GKG05}
S.P.D. {Gill}, A.~{Knebe}, B.K. {Gibson}:
MNRAS \textbf{356}, 1327 (2005)

\bibitem{mamon:BM82}
J.~{Binney}, G.A. {Mamon}:
MNRAS \textbf{200}, 361 (1982)

\bibitem{mamon:CYE97}
R.G. {Carlberg}, H.K.C. {Yee}, E.~{Ellingson}:
ApJ \textbf{478}, 462 (1997)

\bibitem{mamon:NFW96}
J.F. {Navarro}, C.S. {Frenk}, S.D.M. {White}:
ApJ \textbf{462}, 563 (1996)

\bibitem{mamon:LM03}
E.L. {{\L}okas}, G.A. {Mamon}:
MNRAS \textbf{343}, 401 (2003)

\bibitem{mamon:KBM04}
P.~{Katgert}, A.~{Biviano}, A.~{Mazure}:
ApJ \textbf{600}, 657 (2004)

\bibitem{mamon:MG04}
A.~{Mahdavi}, M.J. {Geller}:
ApJ \textbf{607}, 202 (2004)

\bibitem{mamon:Carlberg+01_groups}
R.G. {Carlberg}, H.K.C. {Yee}, S.L. {Morris}, et al.:
ApJ \textbf{552}, 427 (2001)

\bibitem{mamon:LMS03}
Y.T. {Lin}, J.J. {Mohr}, S.A. {Stanford}:
ApJ \textbf{591}, 749 (2003)

\bibitem{mamon:EMN96}
A.E. {Evrard}, C.A. {Metzler}, J.F. {Navarro}:
ApJ \textbf{469}, 494 (1996)

\bibitem{mamon:Sanderson+03}
A.J.R. {Sanderson}, T.J. {Ponman}, A.~{Finoguenov}, E.J. {Lloyd-Davies}, 
  M.~{Markevitch}:
MNRAS \textbf{340}, 989 (2003)

\bibitem{mamon:APP05}
M.~{Arnaud}, E.~{Pointecouteau}, G.W. {Pratt}:
A\&A \textbf{441}, 893 (2005)

\bibitem{mamon:PS97}
P.~{Prugniel}, F.~{Simien}:
A\&A \textbf{321}, 111 (1997)

\bibitem{mamon:ML05b}
G.A. {Mamon}, E.L. {{\L}okas}:
MNRAS \textbf{363}, 705 (2005)

\bibitem{mamon:MB06}
G.A. {Mamon}, G.~{Bou\'e}:
MNRAS to be submitted (2006)

\bibitem{mamon:BG03}
A.~{Biviano}, M.~{Girardi}:
ApJ \textbf{585}, 205 (2003)

\bibitem{mamon:Lokas+06}
E.L. {{\L}okas}, R.~{Wojtak}, S.~{Gottl{\"o}ber}, G.A. {Mamon}, F.~{Prada}:
MNRAS \textbf{367}, 1463 (2006)

\bibitem{mamon:ML05a}
G.A. {Mamon}, E.L. {{\L}okas}:
MNRAS \textbf{362}, 95 (2005)

\bibitem{mamon:JPCS06}
D.H. {Jones}, B.A. {Peterson}, M.~{Colless}, W.~{Saunders}:
MNRAS \textbf{369}, 25 (2006)

\end{thebibliography}
%

\end{document}